\documentclass[mathleft,fleqn,%
]{an}
\def\Mstd{$M_{\mbox{\scriptsize std}}$ }
\def\mins{$m_{\mbox{\scriptsize ins}}$ }
\def\fdif{$f_{\mbox{\scriptsize diff}}$ }
\def\fref{$f_{\mbox{\scriptsize dref}}$ }
\def\sref{$\sigma_{\mbox{\scriptsize ref}}$ }
\def\sdiff{$\sigma_{\mbox{\scriptsize diff}}$ }

\usepackage{graphicx}
\usepackage[varg]{txfonts}
\usepackage{natbib}
\bibpunct{(}{)}{;}{a}{}{,}
\begin{document}

\Pagespan{1}{}
\Yearpublication{2014}%
\Yearsubmission{2014}%
\Month{0}%
\Volume{999}%
\Issue{0}%
\DOI{asna.201400000}%

\title{Metallicity and distance of NGC 6362 from its RR Lyrae and\\ 
SX Phoenicis stars\thanks{Data obtained at the 
CASLEO, Las Campanas and Bosque Alegre Observatories. Complejo Astron\'{o}mico El
Leoncito (CASLEO) is operated under agreement between the Consejo Nacional de
Investigaciones Cient\'{\i}ficas y T\'{e}cnicas de la Rep\'{u}blica Argentina,
and the National Universities of La Plata, C\'{o}rdoba, and San
Juan, Argentina.
}}

\author{A. Arellano Ferro\inst{1}\fnmsep\thanks{Corresponding author:
        {armando@astro.unam.mx}}
J. A. Ahumada\inst{2} , I.H. Bustos Fierro\inst{2}, J. H. Calder\'on\inst{2,3} \and
N. I. Morrell\inst{4}
}
\titlerunning{Metallicity and distance of NGC 6362 from its RR Lyrae and SX
Phoenicis stars}
\authorrunning{Arellano Ferro et al.}
\institute{
Instituto de Astronom\'ia, Universidad Nacional Aut\'onoma de M\'exico, Ciudad
de M\'exico, CP 04510, Mexico
\and 
Observatorio Astron\'omico, Universidad Nacional de C\'ordoba, Laprida 854,
5000 C\'ordoba, Argentina.
\and
Consejo Nacional de Investigaciones Cient\'ificas y T\'ecnicas (CONYCET), Argentina.
\and 
Las Campanas Observatory, Carnegie Observatories, Casilla 601, La Serena, Chile}

\received{XXXX}
\accepted{XXXX}
\publonline{XXXX}


\abstract{%
  New time-series \emph{VI} CCD photometry of the globular cluster NGC 6362 is
studied
with the aim of estimating the reddening, mean metallicity and distance of the cluster
from its population of RR Lyrae stars. The Fourier decomposition of carefully selected
single-mode RR Lyrae light curves, and the use of well-established semi-empirical
calibrations and
revised zero points, lead to the values of [Fe/H]$_{UVES}$  $-1.066\pm0.126$ and
$-1.08\pm0.16$
and the distance
$7.93\pm0.32$ and $8.02\pm0.15$ kpc from the RRab and RRc stars respectively.
The distribution of RR Lyrae stars in the horizontal branch shows a neat segregation
of pulsating modes
about the red edge of the first overtone instability strip, which is not necessarily
expected in an OoI type cluster like NGC 6362.  Four RRab stars are found likely
advanced in their evolution towards the AGB. One new foreground SX Phe star, some 4
kpc in front of the cluster and projected onto the field of our images is reported. We
comment on the heavy light contamination, by a very close neighbouring star, of the
peculiar double-mode V37 variable, recently postulated as a non-typical RRc variable.}

\maketitle

\section{Introduction}
The southern globular cluster NGC~6362 (C1726$-$670  in the
IAU nomenclature), located at $\alpha = 17^\mathrm{h}~31^\mathrm{m}~55.0^\mathrm{s}$, 
$\delta = -67^{\circ}~02^{'}~52^{''}$ (J2000),
 $l = 325.^{\!\!\circ}55$, $b=-17.^{\!\!\circ}57$, 
is at the relatively low distance from the Sun 
of 7.6~kpc,  has a low reddening 
$E(B-V) = 0.09$, and has a low concentration ($c=1.09$,
$\rho_0 = 2.29$) \citep{ha96}. It is also worth
mentioning that recently \cite{Dale14},
 based on HST observations, have identified two
different, spatially mixed populations in the cluster,
apparent in the subgiant and red giant branches  in the
$U$ vs.\ $(U-B)$ diagram (their Fig.~4). \cite{Dale14} also claim that NGC~6362 is
one of the least massive globulars ($M_\mathrm{tot} \sim
 5 \times 10^4 M_\odot$)
where multiple populations have been detected so far.

Given its distance, the cluster 
is bright, with a horizontal branch at $V \sim  15$~mag,
and consequently the discovery of its variable
stars began early. The first 15 variables were found
by \cite{wo19} on plates taken with the 13-inch Boyden telescope 
at Arequipa, Peru.
Much later, \cite{va61} published the discovery of variables
V16 to V31  on plates obtained by P.~Th.~Oosterhoff from South
Africa in 1950 with the 74-inch Radcliffe reflector at Pretoria.
\cite{va61} provided $(x,y)$ coordinates and an ID chart for
all V1-V31 variables. \cite{fou66} found star V32 on plates taken 
with the 60-inch telescope at Bosque Alegre Astrophysical Station, C\'ordoba,
Argentina, again giving coordinates and charts for all the variables discovered
so far. \cite{vh61} discovered V33 (his VH~11) and another seven variables that
had already been independently announced by \cite{va61}. \cite{hsh73} in her
3rd catalogue adopted van~Agt's numbering system, although all the epochs, 
periods, and magnitudes listed 
    for the NGC~6362 variables in that catalogue are from van Hoof's 
    paper. Much more recently, already in the CCD era,
		\cite{maz99}  discovered variables V34--52
		on images taken between 1991 and 1996 with the 2.5-m (du~Pont) 
		and 1-m (Swope) telescopes
		at Las~Campanas, Chile, and gave AR and Dec
		coordinates
    with individual finding charts.
The non-membership status of the eclipsing EC variables V43, V45, and V52
was confirmed by \cite{ruc00}. 
In the Catalogue of Variable Stars in Globular Clusters (CVSGC;
 \citealt{cle01}), the periods, magnitudes, amplitudes, and classifications for V1--37
are from \cite{ole01}, and for V38--52 are from \cite{maz99}, 
while the RA and Dec for most of V1--52 are from \cite{sam09}, with  the
exception of stars V11, 23, 24, 26, 28, 29, 32, and 38--41 which are from
 \cite{maz99}.
Finally, \cite{kal14} reported  a search for variable stars carried out between
1995 and 2009 in the field of NGC~6362 also with the du~Pont and Swope
 telescopes at Las~Campanas; they found 25 newly detected variable stars
(V53--77), including 18 proper-motion cluster members. \cite{kal14} provide
identification
charts and RA and Dec coordinates, although M.~Rozyczka (private comm.)
pointed out that the coordinates of V42, 45, 49, 53--58, 60, 63, and 75--77
 are incorrect in Table~1 of their paper. A new set of high quality data obtained
between 1999 and 2009 was recently employed by Smolec et al. (2017) (hereinafter
Smo17), to discuss the
Blazhko and double mode nature of the sample of RRab and RRc stars. With the
observations reported
in the present work, there is a time base of almost a century of variable star
studies in NGC~6362.

As in most of our recent papers we have employed
the DanDIA\footnote{DanDIA is built from the DanIDL library of IDL routines available
at \texttt{http://www.danidl.co.uk}.} 
 implementation of difference image analysis
(DIA) (\citealt{bra08}) to extract
high-precision photometry for all of the point sources in the
field of NGC~6362. We collected 12245 light curves in the $V$ and $I$
bandpasses with the aim of building up a colour-magnitude
diagram (CMD) and discussing the horizontal branch (HB)
structure as compared to other Oosterhoff type~I (OoI) and
Oosterhoff type II (OoII) clusters\footnote{Oosterhoff (1939) showed that
globular clusters can be distinguished according to the
mean perdiod of their fundamental mode RR Lyrae stars, or RRab, being $\sim$0.55 days
in OoI and $\sim$0.65 days in OoII systems, and noted that the percentage of first
overtone RR Lyrae stars, or RRc, is  lower in OoI than in OoII
clusters.}. We also Fourier decomposed
the light curves of the RR Lyrae stars (RRL) to calculate
their metallicity and luminosity in order to provide independent
and homogeneous estimates of the cluster mean
metallicity and distance.

The scheme of the paper is as follows: In $\S$ 2 we
describe the observations, data reduction and calibration
to the standard system. In $\S$ 3 the periods and phased
light curves of RRL stars are displayed and the Fourier
light curve decomposition of stable RRL is described and 
the reddening is estimated. The
corresponding individual values of [Fe/H] and $M_V$ are reported.
$\S$ 4 summarizes some properties of the SX Phe population in the cluster 
and reports a newly found foreground SX Phe variable. In $\S$ 5
we report the distance to the cluster obtained from the Fourier decomposition 
of RRL stars light curves and the P-L relation of the SX Phe stars.
$\S$ 6 deals with the discussion of the distribution
of RRL in the HB. $\S$ 7 offers a summary of the results. Appendix A gives the
discussion of a few peculiar stars and in Appendix B we display a detailed finding
chart of all variables in the field of NGC 6362.

\begin{figure}
\includegraphics[scale=0.9]{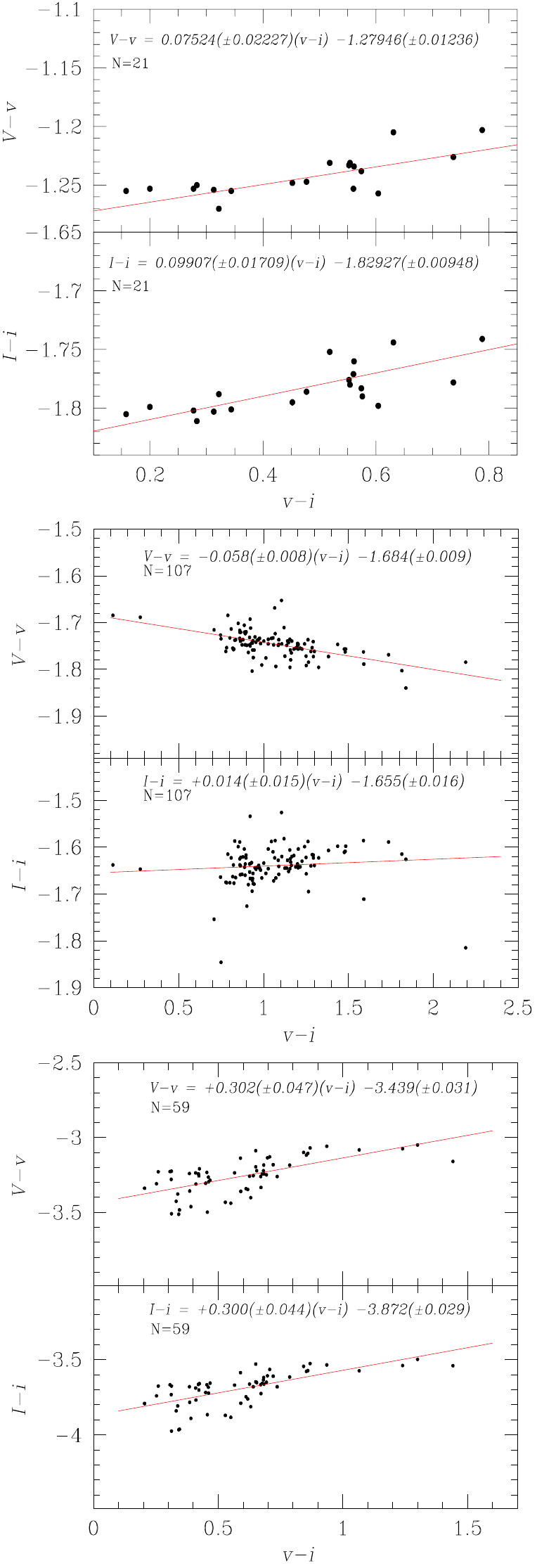}
\caption{The transformation relation for $V$ and $I$ for the three sets of data:
from top to bottom; CASLEO, SWOPE and Bosque Alegre. Note that data from Bosque
Alegre have the largest colour dependence. The transformation equations are given in
each panel.}
\label{Trans}
\end{figure}

\section{Observations}
The observations of this cluster were performed in three sites. 
First, the Complejo Astron´omico El Leoncito (CASLEO),
San Juan, Argentina, where the 2.15-m telescope was used
on March 20--22, 2013. The detector was a Roper Scientific
back-illuminated CCD of 2048$\times$2048 pixels with a scale
of 0.15 arcsec/pix and a field of view (FoV) of approximately 5.1$\times$5.1
arcmin$^2$. Second,the 1-m
Swope Telescope of the Las Campanas Observatory was employed
on June 3 and 14, 2014,  and the
E2V CCD231-84 of 4096x4112 pixels with a scale of 0.435
arcsec/pix and a FoV of approximately 14.5x14.5 arcmin$^2$.
Finally, on June 20, August 21 and 22 and September 5 and 6, 2015,
we used the 1.54-m telescope of the Bosque Alegre Observatory, Cordoba, Argentina
equipped with a CCD Alta U9 of 3072$\times$2048 pixels, with a scale of
0.247 arcsec/pix and a FoV of approximately 12.6$\times$8.4 arcmin$^2$.
The log of our observations is given in Table \ref{tab:observations}.

\subsection{Difference image analysis}
As in previous papers we employed the difference image analysis (DIA) technique and
the DanDIA pipeline (\cite{bra08}; \cite{Bra13}) to extract high-precision
photometry of all point sources in the images of NGC~6362. The procedure and its
caveats have been described in detail by \cite{Bra11}.

\subsection{Photometric calibrations}

\subsubsection{Relative calibration}

Systematic errors in photometric data may be so severe that they mimic bona fide
stellar variability. Time-series photometry of a set of non-variable objects may be
used to calculate and correct these systematic errors. We have applied the methodology
developed by \cite{BF12} to solve for magnitude offsets $Z_k$ that
should be applied to each photometric measurement from image $k$.

\subsubsection{Absolute calibration}

The transformation to the standard \emph{VI} system was performed using the local
standards in the FoV from the collection of
\cite{Stet00}\footnote{
http://www.cadc-ccda.hia-iha.nrc-cnrc.gc.ca/en/community/
STETSON/standards/}. The
calibrations for the three telescopes are shown in Fig. \ref{Trans}. The
transformation equations to the standard $V$ and $I$ system for each data set are
given in the corresponding panel. In Table \ref{tab:vi_phot} we report the $V$ and
$I$ photometry for all RRL stars inour FoV. The full table is published in electronic
format, although we include a small portion of it the printed version.

\begin{table}[t]
\footnotesize
\caption{Observations log of NGC~6362. Data are from three sites;
CASLEO (CAS), Las Campanas (LC) and Bosque Alegre (BA).
Columns $N_{V}$ and $N_{I}$ give the number of images taken with the $V$ and $I$
filters respectively. Columns $t_{V}$ and $t_{I}$ provide the exposure time,
or range of exposure times. In the last column the average seeing is listed.}
\centering
\begin{tabular}{lcccccc}
\hline
Date  & Site &$N_{V}$ & $t_{V}$ (s) & $N_{I}$ &$t_{I}$ (s)&seeing (") \\
\hline
 20130321 &CAS& 15 & 300 & 9 &240  & 2.7\\
 20130322 &CAS& 8 & 480 & 9 & 220  & 2.7\\
 20130323 &CAS& 8 & 360 & 11 & 240  & 2.0\\
 20140504 &CAS& 22 & 480 & 30 & 180-220  & 3.1\\
 20140605 &LC&140 & 4-20 &79 & 2-10 &2.1\\
 20140615 &LC& 88 & 4-20 & 53& 2-10 &1.7\\
 20150621 &BA& 37 & 180-300 & 36 & 120-180 &1.8\\
 20150821 &BA& 8 & 300 & 11 & 150 &3.5\\
 20150821 &BA& 11 & 300  & 14 & 150 &2.7\\
 20150906 &BA& 27 & 300 & 27 & 150 &2.5  \\
 20150907 &BA& 3 & 300 & 6 & 150& 1.9 \\
\hline
Total:   && 367&  -- & 285 & -- & --\\
\hline
\end{tabular}
\label{tab:observations}
\end{table}

\begin{table*}
\footnotesize
\caption{Time-series \textit{V} and \textit{I} photometry for all RR Lyrae stars in our 
field of view. The standard \Mstd and
instrumental \mins magnitudes are listed in columns 4 and~5,
respectively, corresponding to the variable stars in column~1. Filter and epoch of
mid-exposure (in heliocentric julian days) are listed in columns 2 and 3, respectively. The uncertainty on
\mins is listed in column~6, which also corresponds to the
uncertainty on \Mstd. For completeness, we also list the
reference and differential fluxes \fref 
and \fdif and the scale factor \lowercase{\textit{p}} 
in columns 7, 9, and~11, along with the uncertainties \sref 
and \sdiff in columns 8 and~10. This is an extract from
the full table, which is available with the electronic version of the article.}
\centering
\begin{tabular}{ccccccccccc}
\hline
Variable &Filter & HJD & $M_{\mbox{\scriptsize std}}$ &
$m_{\mbox{\scriptsize ins}}$
& $\sigma_{m}$ & $f_{\mbox{\scriptsize ref}}$ & $\sigma_{\mbox{\scriptsize ref}}$ &
$f_{\mbox{\scriptsize diff}}$ &
$\sigma_{\mbox{\scriptsize diff}}$ & $p$ \\
Star ID  &    & (d) & (mag)     & (mag)   & (mag) & (ADU s$^{-1}$) &(ADU s$^{-1}$)   
               &(ADU s$^{-1}$)  &(ADU s$^{-1}$)    & \\
\hline
V1 &$V$ &2456372.79007& 14.994& 16.275&  0.001 &  631.204 & 1.835 & $+$536.163&3.076&3.7921\\
V1 &$V$ &2456372.79452& 15.015& 16.296&  0.001 &  631.204 & 1.835 & $+$481.905&2.989&3.8046\\
\vdots   &  \vdots  & \vdots & \vdots & \vdots & \vdots   & \vdots & \vdots  & \vdots&\vdots \\
V1 &$I$&2456372.78347 & 14.559 & 16.390& 0.003&   1406.814&  3.984& $-$33.805&  7.146& 1.9506\\
V1 &$I$&2456372.80310 & 14.602 & 16.434& 0.002&   1406.814&  3.984& $-$140.873& 3.975& 1.9684\\
\vdots   &  \vdots  & \vdots & \vdots & \vdots & \vdots   & \vdots & \vdots  & \vdots&\vdots \\
V2 &$V$&2456372.79007&15.040&16.317& 0.001& 489.291& 1.760& $+$963.456 &2.956&3.7921\\
V2 &$V$&2456372.79452&15.060&16.337& 0.001& 489.291& 1.760& $+$914.472 &2.881&3.8046\\
\vdots   &  \vdots  & \vdots & \vdots & \vdots & \vdots   & \vdots & \vdots  & \vdots&\vdots \\
V2 &$I$&2456372.78347 & 14.583& 16.409&  0.003&1194.571&   4.129 &$+$333.684 &6.998& 1.9506\\
V2 &$I$&2456372.80310 & 14.639& 16.466&  0.002&1194.571&   4.129 &$+$200.011 &3.851& 1.9684\\
\vdots   &  \vdots  & \vdots & \vdots & \vdots & \vdots   & \vdots & \vdots  & \vdots&\vdots \\
\hline
\end{tabular}
\label{tab:vi_phot}
\end{table*}

\section{The RR Lyrae stars}

\begin{figure*}
\includegraphics[scale=0.90]{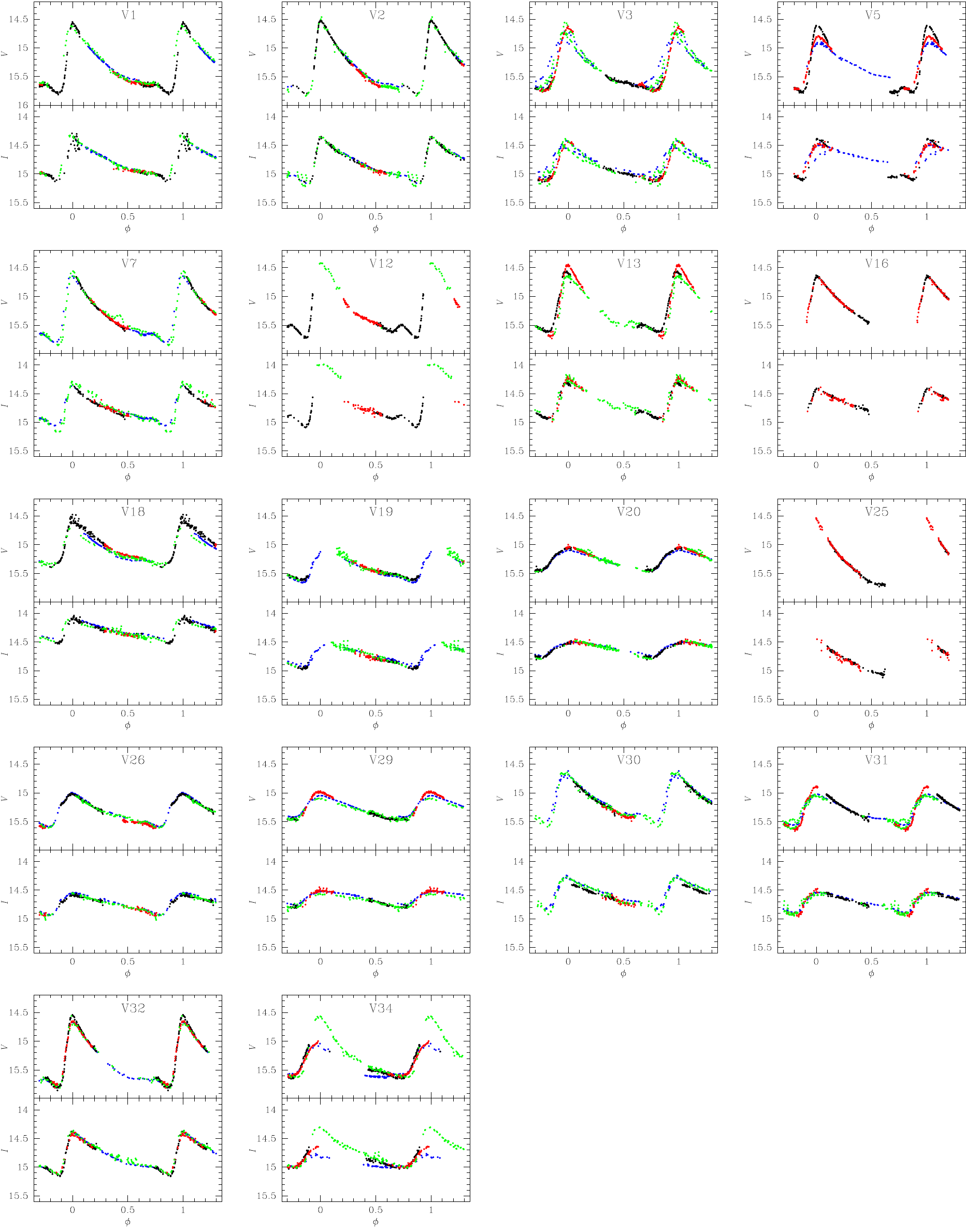}
\caption{Light curves of RRab stars. Colours are assigned as per observatory as
follows: blue for CASLEO, red and black for Las Campanas and green for Bosque Alegre.}
\label{RRab}
\end{figure*}

\begin{figure*}
\includegraphics[scale=0.90]{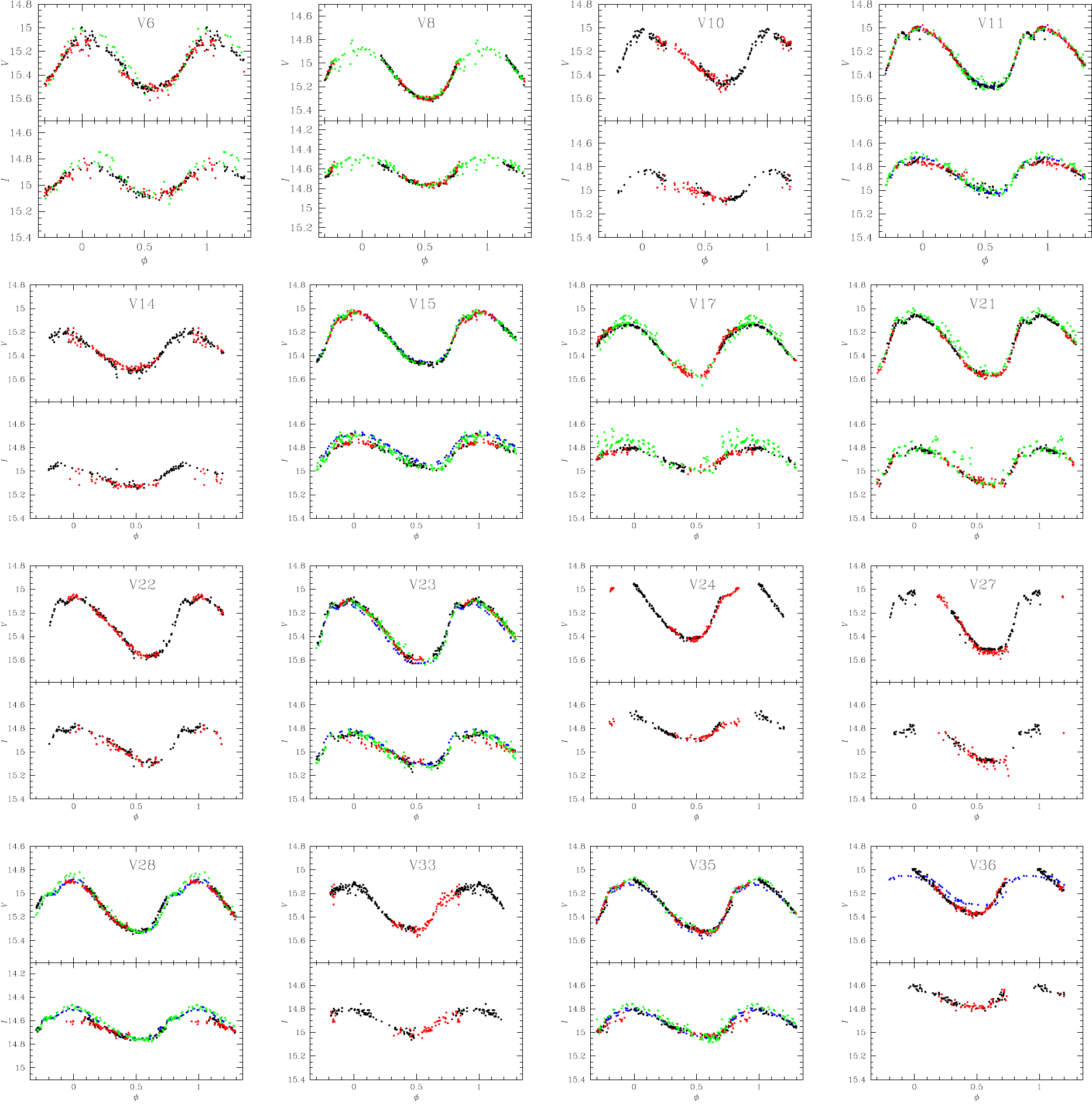}
\caption{Light curves of RRc stars. Colours are as in Fig. \ref{RRab}.}
\label{RRc}
\end{figure*}

The resulting light curves of  the RR Lyrae stars combining the observations from 
the runs in the three sites are shown in  Figs. \ref{RRab} and \ref{RRc}. The zero
points for CASLEO and LC match very well. For BA however we found small drifts
that vary from star to star. With the aim of using all data to refine the
period, we applied these drifts and then the light curves were phased with the new
resulting periods listed in column 9 of Table \ref{variables}. We also 
report itensity-weighted mean $V$ and $I$
magnitudes, amplitudes and equatorial coordinates for all the RR Lyrae stars.
For comparison we also include the periods recently reported by Smo17.

\begin{table*}
\begin{center}

\caption{General data for the RR Lyrae stars in NGC~6362 in the FoV of 
our images. The variable types are adopted from the work of Smo17. Previous period estimates for each variable from Smo17 are reported in
column~7 for comparison with our periods in column~9.}
\label{variables}

\begin{tabular}{llllllllllll}
\hline
Variable & Variable & $<V>$ & $<I>$   & $A_V$  & $A_I$   & $P$ (Smo17) & 
HJD$_{\mathrm{max}}$ 
& $P$ (this work)    & RA   & Dec         \\
Star ID  & Type     & (mag) & (mag)   & (mag)  & (mag)   & (d)  & 
($+2\,450\,000$) & (d) & (J2000.0)   & (J2000.0)    \\
&&&&&&&&&&&\\
\hline
V1  & RRab  \emph{Bl} & 15.352 & 14.804 & 1.292 & 0.782 & 0.50479162 & 6813.9219&0.504814 & 17:31:54.72& $-$67:02:45.7\\
V2  & RRab            & 15.368 & 14.834 & 1.366 & 0.856& 0.48897301 & 6813.7537 & 0.488972 & 17:31:50.23 & $-$67:04:25.4\\
V3  & RRd             & 15.378 & 14.866 & 1.185 & 0.774 & 0.44728792 & 6823.8616 & 0.447297 & 17:31:40.91 & $-$67:04:15.8 \\
V5  & RRab \emph{Bl}  & 15.348 & 14.761 & 1.177 & 0.716 & 0.52083783 & 6813.8986 & 0.521434 &17:32:08.82 & $-$67:02:59.4 \\
V6  & RRc  \emph{Bl}  & 15.312 & 14.911 & 0.47  & 0.28  & 0.26270671 & 6813.8662 & 0.238084 & 17:32:03.85 & $-$66:59:51.8 \\
V7  & RRab \emph{Bl}  & 15.368 & 14.783 & 1.242 & 0.872 & 0.521581388 & 6781.8912 & 0.521572 &17:31:58.52 & $-$67:01:01.4 \\
V8  & RRc \emph{nr}   & 15.082 & 14.620 & 0.517 & 0.321 & 0.38148471 & 7194.7619 & 0.381466 & 17:31:10.05 & $-$67:01:01.3 \\ 
V10 & RRc \emph{Bl}   & 15.257 & 14.970 & 0.471 & 0.254 & 0.265638816 & 6813.8947 & 0.315140 & 17:32:26.13 & $-$66:56:52.6 \\
V11 & RRc             & 15.255 & 14.869 & 0.502 & 0.317 & 0.288789268 & 6823.7797 & 0.288789 & 17:31:49.90 & $-$67:01:57.8 \\
V12 & RRab \emph{Bl}  & 15.227 & 14.641 & 1.308 & 1.103 & 0.5328814 & 7194.6017 & 0.533046 & 17:31:13.13 & $-$67:04:30.9 \\
V13 & RRab \emph{Bl}  & 15.258 & 14.651 & 1.251 & 0.741 & 0.58002740 & 6823.8027 & 0.580010 & 17:31:15.14 & $-$67:04:47.6 \\
V14 & RRc             & 15.374 & 15.056 & 0.344 & 0.213 & 0.24620647 & 6823.7091 & 0.252368 & 17:32:57.94 & $-$67:02:12.8 \\
V15 & RRc \emph{nr}   & 15.263 & 14.841 & 0.448 & 0.290 & 0.279945707 & 6823.7574 & 0.279944& 17:32:03.47 & $-$67:02:44.3 \\
V16 & RRab            & 15.359 &14.774  &1.12   & 0.67  & 0.525674215 & 6813.7472 & 0.525711 &17:31:58.10&$-$67:07:12.0 \\
V17 & RRc  \emph{nr}  & 15.335 & 14.861 & 0.511 & 0.265 & 0.31460473 & 6813.8251 & 0.314604 & 17:32:29.46 & $-$67:03:51.2 \\
V18 & RRab \emph{Bl}  & 15.073 & 14.320 & 0.913 & 0.461 & 0.51288484 & 6813.7955 & 0.512899 & 17:32:13.59 & $-$67:01:33.1 \\
V19 & RRab            &15.365  & 14.730 & 0.663 & 0.480 & 0.59450528 & 6781.9170 &0.594511   &17:32:15.96 &$-$67:03:09.4  \\
V20 & RRab \emph{Bl}  & 15.273 & 14.627 & 0.436 & 0.329 & 0.69835898 & 6823.7125 & 0.698361 &17:32:02.64 & $-$67:02:59.4 \\
V21 & RRc  \emph{nr}  & 15.321 & 14.945 & 0.535 & 0.362 & 0.281390043 & 6813.7955 &0.281392 & 17:32:22.59 & $-$67:04:30.6 \\
V22 & RRc             & 15.318 & 14.939 & 0.516 & 0.335 & 0.26683523 & 6823.7091 & 0.253262 &17:32:26.74 & $-$67:07:55.1 \\
V23 & RRc  \emph{nr}  & 15.359 & 14.980 & 0.571 & 0.345 & 0.275105063 & 6813.8251 & 0.275108 &17:32:00.12 & $-$67:03:07.5 \\
V24 & RRc  \emph{nr}  & 15.195 & 14.795 & 0.489 & 0.240 & 0.32936190 & 6813.7257 &0.329318  &17:32:07.16&$-$67:03:20.3  \\
V25 & RRab            &15.45   & 14.9   &1.23   &0.74   & 0.455890887 & 6823.6985 &0.455824 &17:30:54.37&$-$67:06:18.5\\
V26 & RRab            & 15.351 & 14.750 & 0.610 & 0.403 & 0.60217449 & 6781.8739 & 0.602179 &17:31:58.87 & $-$67:03:22.2 \\ 
V27 & RRc   \emph{nr} & 15.253 & 14.957 & 0.520 & 0.303 & 0.27812399 & 6813.9520 & 0.322892 &17:31:21.62 & $-$66:56:27.7 \\
V28 & RRc  \emph{Bl?} & 15.111 & 14.624 & 0.457 & 0.271 & 0.3584133 & 6372.8559 &0.358399  &17:31:59.16  & $-$67:02:07.7 \\
V29 & RRab \emph{Bl}  & 15.272 & 14.658 & 0.502 & 0.321 & 0.64778329 & 6823.8064 & 0.647743 &17:31:52.49 & $-$67:03:20.2 \\
V30 & RRab \emph{Bl}  & 15.186 & 14.599 & 0.954 & 0.642 & 0.61340457 & 6373.8773 & 0.613415 &17:31:39.59 & $-$67:01:33.6 \\
V31 & RRab \emph{Bl}  & 15.323 & 14.714 & 0.749 & 0.460 & 0.60021294 & 6823.8726 & 0.600220 &17:31:49.20 & $-$67:01:20.7 \\
V32 & RRab \emph{Bl}  & 15.385 & 14.817 & 1.258 & 0.765 & 0.49724171 & 6813.8468 & 0.497248 &17:32:01.83 & $-$67:02:12.7 \\
V33 & RRc  \emph{nr}  & 15.319 & 14.924 & 0.398 & 0.231 & 0.30641758 & 6813.7889 & 0.338096 &17:32:47.91 & $-$66:56:36.2 \\
V34 & RRd             & 15.305 & 14.779 & 1.070 & 0.708 & 0.49432939  &7194.6212  &0.494306 &17:31:52.81 & $-$67:03:34.3 \\
V35 & RRc  \emph{nr}  & 15.318 & 14.883 & 0.462 & 0.289 & 0.29079074 & 6813.7263 & 0.290792 &17:32:08.14 & $-$67:03:03.0 \\
V36 & RRc \emph{Bl/nr} &15.173 & 14.698 &0.393  & 0.248 & 0.31009148  & 6813.7294 & 0.3101   &17:31:43.57 & $-$67:02:16.8 \\
V37 & blRR  ?         & 15.319 & 14.974 & 0.403 & 0.176 & 0.25503903  & 6813.9287 & 0.254576 &17:31:32.12 & $-$67:02:03.4 \\
\hline
\end{tabular}
\raggedright
\center{\quad \emph{Bl}: Blazhko modulations; \emph{nr}:doble mode with at least a non-radial mode according to Smo17.}
\end{center}
\end{table*}

\subsection{The reddening of NGC~6362 from its RRab stars}
\label{RRab_reddening}

Using the fact that RRab stars have nearly the same intrinsic colour $(B-V)_0$ at
minimum light \citep{Stu66}, one can calculate the individual reddenings of these
stars, which can provide a good average of the cluster reddening or to reveal the
presence of differential reddening. The $(V-I)_0$ at minimum has
been calibrated by \cite{Gul05} as $(V-I)_{0,min} = 0.58 \pm 0.02$. We
have adopted this value and the minimum $(V-I)$ from our observations to estimate
$E(V-I)$ for each
RRab with a well defined minimum. This was converted to $E(B-V)$ through the ratio 
$E(V-I)/E(B-V)= 1.259$ derived from \cite{Sch98}.
From 15 RRab stars we find an average of $E(B-V)=0.063 \pm 0.024$. The scatter is
small
and shows no signs of differential reddening, hence, we shall adopt this value for the
determination of the distance to NGC 6362 via the several approaches described in
$\S$ \ref{Distance}. The above reddening calculation can be compared with previous
estimations such as 0.10 \citep{ole01}, 0.06$\pm$0.03 \citep{Piot99} and 0.08
\citep{Broc99}.

\subsection{Fourier decomposition and physical parameters of RR Lyrae stars}

Determination of [Fe/H] and $M_V$ of RR Lyrae stars in a given cluster enables the
estimation of the mean values of the metallicity and distance of the parental
cluster.
This can be achieved via the Fourier
decomposition of their light curves and the employment of well established
calibrations and their zero points of these physical quantities and the corresponding
Fourier
parameters. By doing it on a homogeneous basis, i.e. using the same semi-empirical
calibrations and zero points for a family of globular clusters of both Oosterhoff
types, OoI and OoII, an independent insight on the metallicity dependence of
the horizontal branch (HB) luminosity, i.e.,the familiar $M_V$-[Fe/H] relation can
be obtained. 
Such approach has been applied by \cite{Are17} to a group of 23 globular clusters.
While preliminary results for NGC 6362 were included in that paper, in the present
work we publish the specific values of the Fourier parameters and the individual
physical parameters for a carefully selected sample of stable RRab stars and
RRc stars in the cluster, according to the accurate photometry of Smo17.

\begin{table*}
\begin{center}

\caption{Fourier coefficients of RRab and RRc stars in NGC~6362. The numbers in 
parentheses indicate the uncertainty on the last decimal place. Also listed are the 
number of harmonics~$N$ used to fit the light curve of each variable and the
deviation 
parameter~$D_m$.}     

\label{tab:fourier_coeffs} 
\begin{tabular}{lllllllllrr}
\hline
Variable     & $A_{0}$    & $A_{1}$   & $A_{2}$   & $A_{3}$   & $A_{4}$  
&$\phi_{21}$
& $\phi_{31}$ & $\phi_{41}$ 
& $N$   &$D_m$ \\
ID     & ($V$ mag)  & ($V$ mag)  &  ($V$ mag) & ($V$ mag)& ($V$ mag) & & &  & & \\
\hline
\multicolumn{11}{c}{RRab} \\
\hline
V2  & 15.368(2) & 0.4145(3) & 0.2088(3) & 0.1458(3) & 0.0974(3) & 3.888(2) & 8.147(3) &6.034(4)  & 10 & 1.7\\
V16 & 15.359(5) & 0.3442(3) & 0.1796(3) & 0.1212(3) & 0.0816(3) & 3.973(2) & 8.234(4) &6.239(4)  & 10 & 0.9\\
V19 & 15.365(2) & 0.2080(2) & 0.0945(2) & 0.0542(2) & 0.0235(2) & 4.172(3) & 8.656(4) &7.097(9)  & 10 & 1.0\\
V25 & 15.263(2) & 0.4195(4) & 0.2018(4) & 0.1510(4) & 0.0966(4) & 3.831(2) & 7.975(4) &5.871(5)  & 10 & 1.4 \\
V26 & 15.351(1) & 0.2131(2) & 0.0973(2) & 0.0555(2) & 0.0245(2) & 4.200(3) & 8.690(5) &7.139(9)  & 10 & 1.3\\
\hline
\multicolumn{11}{c}{RRc} \\
\hline
V11 & 15.255(2) & 0.2434(2) & 0.0333(2) & 0.0207(2) & 0.0167(2) & 4.830(8) & 3.410(12) & 2.140(15) & 9 & \\
V14 & 15.374(1) & 0.1587(4) & 0.0184(4) & 0.0040(4) & 0.0014(4) & 4.512(22)& 2.905(99)& 1.092(281) & 9 & \\
V15 & 15.263(1) & 0.2192(2) & 0.0306(2) & 0.0190(2) & 0.0131(2) & 4.593(6) & 3.098(10) & 2.002(15) & 9 & \\
V21 & 15.321(1) & 0.2366(2) & 0.0331(2) & 0.0222(2) & 0.0159(2) & 4.705(6)  & 3.237(8) & 2.084(12) & 9 & \\
V22 & 15.318(1) & 0.2303(8) & 0.0379(8) & 0.0148(8) & 0.0134(8) & 4.680(21) & 2.850(52)& 1.588(58) & 9 & \\
V23 & 15.359(1) & 0.2428(2) & 0.0377(2) & 0.0232(2) & 0.0170(2) & 4.651(6)  & 2.907(8) & 1.770(10) & 9 & \\
\hline
\end{tabular}
\end{center}
\end{table*}

\begin{table*}
\begin{center}

\caption{Physical parameters of the stable RRab and RRc stars. The numbers in parentheses 
indicate the uncertainty on the last decimal places and have been calculated as 
described in the text.} 

\label{fisicos}

 \begin{tabular}{lccccccc}
\hline 
Star&[Fe/H]$_{ZW}$ & [Fe/H]$_{UVES}$ &$M_V$ & log~$T_{\rm eff}$  &log$(L/{\rm
L_{\odot}})$ &$M/{\rm M_{\odot}}$&$R/{\rm R_{\odot}}$\\
\hline
\multicolumn{8}{c}{RRab} \\
\hline
V2 &-1.274(3)& -1.157(3)& 0.633(1)& 3.821(7)& 1.647(1)&0.69(6)&5.09(1)\\
V16&-0.986(4)& -0.887(4)& 0.789(1)& 3.826(7)& 1.584(1)&0.65(5)&4.63(1)\\
V19&-1.194(4)& -1.075(3)& 0.637(1)& 3.806(7)& 1.645(1)&0.62(5)&5.46(1)\\
V25&-1.311(4)& -1.195(4)& 0.688(1)& 3.823(7)& 1.625(1)&0.71(6)&4.93(1)\\
V26&-1.190(5)& -1.211(32)& 0.610(3)& 3.824(10)& 1.656(1)& 0.51(6)& 5.09(1)\\
 
\hline
Weighted mean& $-$1.203(2)&$-$1.066(2)& 0.657(1)& 3.816(3)& 1.637(1)&0.66(2)&5.22(1)\\
$\sigma$&$\pm$0.126&$\pm$0.126&$\pm$0.072&$\pm$0.007&$\pm$0.025&$\pm$0.07&$\pm$0.27\\
\hline
\multicolumn{8}{c}{RRc} \\
\hline
V11&-1.15(2)&-1.03(2) &0.566(1) &3.872(1)&1.674(1)&0.56(1)& 4.16(1) \\
V14&-0.94(16)&-0.85(11)&0.683(2)&3.878(1)&1.627(1)&0.68(2)& 3.84(4)\\
V15&-1.16(2)&-1.11(2) &0.601(1) &3.872(1)&1.660(1)&0.60(1)& 4.09(1)\\
V21&-1.16(1)&-1.04(1) &0.582(1) &3.873(1)&1.667(1)&0.58(1)& 4.11(1)\\
V22&-1.00(9)&-0.90(6) &0.621(4) &3.877(1)&1.652(1)&0.62(4)& 3.96(2)\\ 
V23&-1.28(14)&-1.16(1)&0.585(1) &3.872(1)&1.666(1)&0.59(1)& 4.12(1)\\
\hline
Weighted mean&
$-$1.21(1)&$-$1.08(1)&0.589(1)&3.872(1)&1.664(1)&0.59(1)&4.10(1)\\
$\sigma$&$\pm$0.16&$\pm$0.16&$\pm$0.046&$\pm$0.003&$\pm$0.015&$\pm$0.04&$\pm$0.11\\
\hline
\end{tabular}
\end{center}
\end{table*}

Although the procedure and the employed calibrations have been described in detail by
\cite{Are17} and in several papers cited there, for completeness we
include here the fundamentals.

The form of the Fourier representation of a given light curve is: 
\begin{equation}
\label{eq.Foufit}
m(t) = A_0 + \sum_{k=1}^{N}{A_k \cos\ ({2\pi \over P}~k~(t-E) + \phi_k) },
\end{equation}

\noindent
where $m(t)$ is the magnitude at time $t$, $P$ is the period, and $E$ is the epoch. A
linear
minimization routine is used to derive the best fit values of the 
amplitudes $A_k$ and phases $\phi_k$ of the sinusoidal components. 
From the amplitudes and phases of the harmonics in Eq.~\ref{eq.Foufit}, the 
Fourier parameters, defined as $\phi_{ij} = j\phi_{i} - i\phi_{j}$, and $R_{ij} =
A_{i}/A_{j}$, 
are computed. 

Since some of the calibrations and zero points employed in this work towards the calculation of stellar physical quantities differ from the ones used in the previous 
Fourier decomposition analysis for the RR Lyrae stars in NGC 6362 \cite{ole01}, 
below we explicitly list the calibrations we used:

\begin{equation}
{\rm [Fe/H]}_{J} = -5.038 ~-~ 5.394~P ~+~ 1.345~\phi^{(s)}_{31},
\label{eq:JK96}
\end{equation}

\begin{equation}
M_V = ~-1.876~\log~P ~-1.158~A_1 ~+0.821~A_3 + K,
\label{eq:ḰW01}
\end{equation}

\noindent given by \cite{JK96} and \cite{KW01},
respectively.  The standard deviations of the above calibrations are 0.14 dex 
\citep{Jur98} and 0.04 mag, respectively. In eq. \ref{eq:ḰW01} we have used K=0.41
(see the discussion in Section~4.2 of \cite{AGB10}).
Eq. \ref{eq:JK96} is applicable to RRab stars with a {\it deviation parameter} $D_m$,
defined by \cite{JK96} and \cite{KK98}, not exceeding an
upper limit. These authors suggest $D_m \leq 3.0$. The $D_m$ is listed in
column 11 of Table~\ref{tab:fourier_coeffs}, where it is obvious that the five 
stable RRab stars have light curves consistent with the calibration of eq.
\ref{eq:JK96}.

For the RRc stars we employ the calibrations: 

$$ {\rm [Fe/H]}_{ZW} = 52.466~P^2 ~-~ 30.075~P ~+~ 0.131~\phi^{(c)~2}_{31}  $$
\begin{equation}
~~~~~~~	~-~ 0.982 ~ \phi^{(c)}_{31} ~-~ 4.198~\phi^{(c)}_{31}~P ~+~ 2.424,
\label{eq:Morgan07}
\end{equation}

\begin{equation}
M_V = 1.061 ~-~ 0.961~P ~-~ 0.044~\phi^{(s)}_{21} ~-~ 4.447~A_4, 
\label{eq:K98}	
\end{equation}

\noindent given by  \cite{Mor07} and \cite{Kov98}, respectively. For eq.
\ref{eq:K98} the zero point was reduced to 1.061 mag to make the
luminosities of the RRc consistent with the distance modulus of 18.5~mag for the LMC (see
discussions by \cite{Cac05} and \cite{AGB10}). The original zero
point given by \cite{Kov98} is 1.261.

When necessary, the coefficients were transformed from 
cosine series phases into the sine series via the relation ~~
$\phi^{(s)}_{jk} = \phi^{(c)}_{jk} - (j - k) {\pi \over 2}$

The values of $A_0$, in $V$ and $I$, for all the RRL stars in NGC 6362 are given in
Table~\ref{variables} as the intensity weighted quantities $<V>$ and
$<I>$. However, several of these stars have been identified as Blazhko variables or
double mode pulsators (Smo17), hence we have limited the Fourier
decomposition, for the purpose of physical parameters determination, to those stars
proven to be stable
in the time scale of the analysis of Smo17. 
Clearly, the light curves of Smo17 are more dense and much better
covered than ours in Figs. \ref{RRab} and \ref{RRc}, since their time-base is much
longer. Hence, we decided to Fourier decompose the light curves from Smo17 for the
rather stable stars (i.e. the RRc stars V11, V14, V15, V21, V22 and
V23, and the RRab stars V2, V16, V19, V25 and V26). The Fourier decomposition parameters for these stars and their corresponding  physical parameters are
listed in Tables \ref{tab:fourier_coeffs} and \ref{fisicos} respectively. 
The absolute magnitude $M_V$ was converted into luminosity with
$\log (L/{\rm L_{\odot}})=-0.4\, (M_V-M^\odot_{\rm bol}+BC$). The bolometric
correction was
calculated using the formula $BC= 0.06\, {\rm [Fe/H]}_{ZW}+0.06$ given by \cite{SC90}. We adopted  $M^\odot_{\rm bol}=4.75$~mag. For the distance
calculation, and given that there are no signs of differential reddening for
NGC~6362, we have adopted $E(B-V)=0.063$ (see $\S$ \ref{RRab_reddening})
The weighted average of [Fe/H] and distance are considered good mean values for the parental cluster. 
We found [Fe/H]$_{ZW} = -1.203\pm 0.126$ and $-1.21\pm 0.16$ for the RRab and RRc
stars respectively,
which in the scale of \cite{Carr09} are [Fe/H]$_{UVES}= -1.066\pm 0.126$ and $-1.08\pm
0.16$.

The value of [Fe/H]$_{UVES}$ obtained above can be compared with independent determinations 
of the metallicity of the cluster that can be found in the literature. The value given in the
catalogue of globular cluster parameters of \cite{ha96} is [Fe/H]$_{UVES}=-0.99$; 
\cite{ole01} finds [Fe/H]$_{ZW}=-1.08$
or [Fe/H]$_{UVES}=-0.97$. On spectroscopic grounds \cite{CG97} found
[Fe/H]=$-0.96$ from an analysis of echelle spectra of cluster
giants, and \cite{Rut97} found [Fe/H]=$-0.99 \pm0.03$ from Ca II
triplet index. More recent spectroscopic determinations from ESO/FLAMES spectra
were obtained by \cite{Mucc16} 
and \cite{Mass17} and obtained  [Fe/H]= $-1.09 \pm 0.01$ and $-1.07 \pm 0.01$
respectively. Thus, our results are in very good agreement with recent 
spectroscopic estimations.

\subsection{The Color-Magnitude Diagram} 

\begin{figure*}
\includegraphics[scale=1.5]{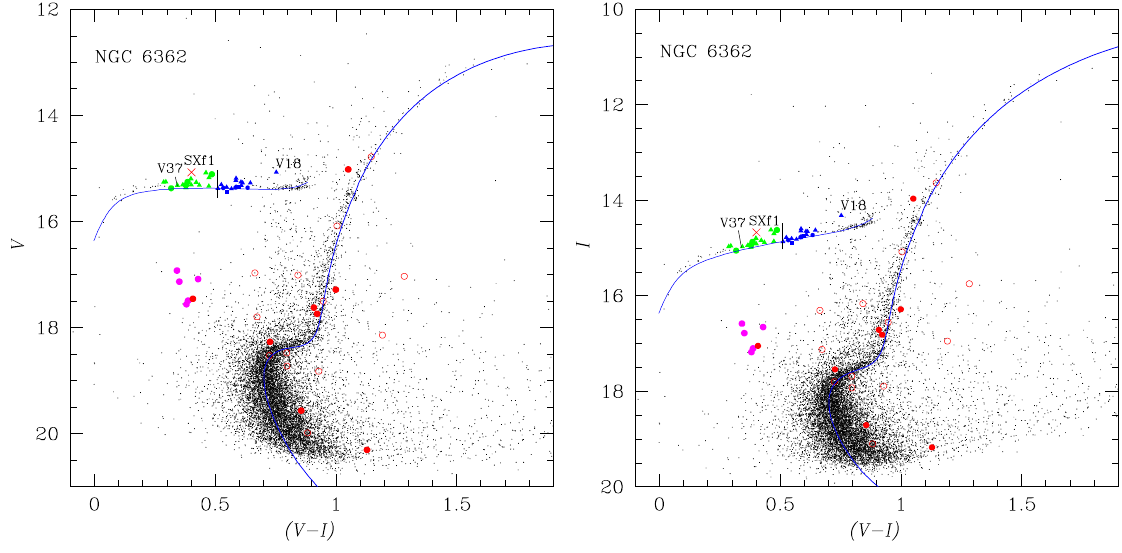}
\caption{CMD of NGC 6362. Coloured symbols are used for variable stars according to
the following code: RRab (blue), RRc (green), SX Phe (purple), binaries in the field of the 
cluster discussed by \cite{kal14} that are members (filled red circles) and non-members (open red circles). 
The red cross shows the position of the newly found variable SXf1 of the SX Phe type. 
See appendix \ref{individual} for a discussion of the labelled stars. The isochrone
for
12.1 Gyr, [Fe/H]=-1.0, Y=0.25 and [$\alpha$/H]=0.4 and the slightly more metal poor
zahb (-1.31) are from
from \cite{vdB14}, and have been reddened by $E(B-V)=0.063$ and shifted to
the average distance modulus found from the RRL stars. The vertical black line
shows the first overtone red edge calculated by \cite{Are16} and
shows the neat mode segregation.}
\label{CMD_6362}
\end{figure*}

The color-magnitude diagram (CMD) in Fig. \ref{CMD_6362} was built using the magnitude
weighted means
of $V$ and $V-I$ for the 12245 stars in the field or our reference image from the run
in Las Campanas. Variable stars are labelled and plotted using their
intensity-weighted
means $<V>$ and $<V>-<I>$ calculated from the light curves in Figs. \ref{RRab} and
\ref{RRc} and listed in Table \ref{variables}. All symbols and colours are explained in 
the caption of the figure. For the age of the cluster we adopted 12.1 Gyr from the differential 
age determination of \cite{dAng05}. The isochrone
and zahb model are from \cite{vdB14} and are shifted to the average distance modulus
found from the RRL stars.

\subsection{Bailey diagram and Oosterhoff type}
\label{secBailey}

\begin{figure}
\includegraphics[scale=0.65]{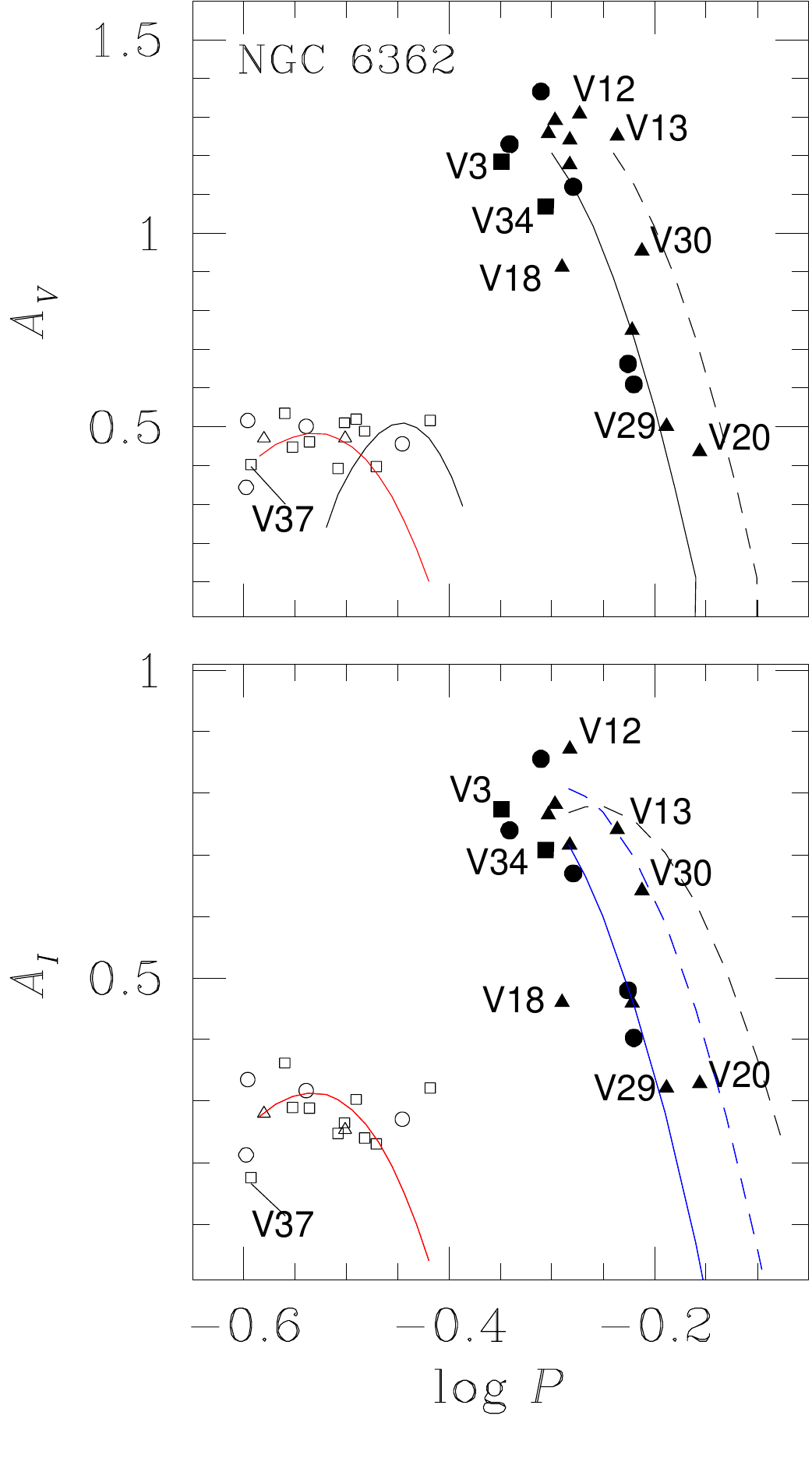}
\caption{Period-Amplitude diagram for NGC 6362. Filled and open symbols represent RRab
and
RRc stars, respectively. Triangles and squares are used for stars with
Blazhko modulations or double mode stars. The continuous
and segmented lines in the top panel are the loci for evolved and unevolved
stars in M3 according to \cite{Cac05}.
The black parabola was obtained by \cite{Kun13a}
from 14 OoII clusters. The red parabolas were calculated
by \cite{Are15} from a sample of
RRc stars in five OoI clusters and avoiding Blazhko variables.
In the bottom panel the black segmented locus
was found by \cite{Are11} and \cite{Are13} for the
OoII clusters NGC 5024 and NGC 6333 respectively. The blue loci
are from \cite{Kun13b}. See $\S$ \ref{secBailey} for a discussion.}
\label{figBailey}
\end{figure}

The log P vs Amplitude diagram, also known as the Bailey diagram, is  a useful tool 
because it is a good discriminator between OoI and OoII clusters, the 
RRab and RRc stars are well separated and hence their classification is confirmed,
and RRab stars of advanced evolution can be identified. Fig. \ref{figBailey} shows the
diagram resulting from 
the data listed in Table \ref{variables}. The full amplitudes were measured on 
the light curves of Figs. \ref{RRab} and \ref{RRc}, except in those cases were we
missed the maximum or minimum. In these cases we estimated the full amplitude from the
curves of Smo17. When amplitude modulations are
evident, the maximum observed amplitude was taken. 
The distribution of the RR Lyrae stars in the diagram confirms NGC 6362 to be of the
OoI type. Some 
outstanding stars deserve a comment. According to \cite{Cac05},
RRab stars closer to the dashed sequence are more evolved. V12, V13, V20 and
V30 
are therefore good candidates to be ahead in their evolution towards the AGB. These
four stars have Blazhko modulations (Smo17). In the amplification of the HB star in
Fig.\ref{HB} they are among the most luminous RRab stars. Conversely the luminous V18
and V29 do not show signs of being more evolved.

\section{The SX Phoenicis stars in NGC 6362}

There are six SX Phe stars identified in NGC 6362. They are listed in Table
\ref{SXPhe} along with their intensity weighted magnitude, periods, pulsating mode, 
equatorial coordinates and the discovering paper. The pulsating modes 
were assigned from the position of the star on the log P- $V$ and the P-L relation of
\cite{CS12} shifted to the predicted distance.

We have identified a new SX Phe in the field of our images in 
NGC 6362. In the CMD the star falls in an odd position for a member SX Phe (red cross
in Fig. \ref{CMD_6362}). The star is at least 2 magnitudes brighter than the SX Phe in
the cluster, thus it must be a foreground object. Therefore, we refrain from assigning
a sequential variable number to the star and 
identify it as SXf1. The light curves in our $V$ and $I$ photometry phased with a
period of 0.07949d is shown in Fig. \ref{SXf1} in appendix \ref{individual}.

\begin{table*}
\begin{center}

\caption{SX Phe stars in the field of NGC 6362.}
\label{SXPhe}

\begin{tabular}{llllllllll}
\hline
Variable &  $<V>$ & $<I>$   & $P$ (Kal14) &  $P$ (this work)&Pulsating &RA   & Dec& Ref.     \\
Star ID  &  (mag) & (mag)   & (d)  &  (d) &mode &(J2000.0)   & (J2000.0)&    \\
&&&&&&&&&\\
\hline
V38  & 16.926 & 16.584 &0.06661582  &0.06661901&F&17:31:43.7  &$-$67:02:58.0&1\\
V46  & 17.491 & 17.103 &0.050634688 &0.05063447&F& 17:32:25.0 &$-$67:00:31.4&1\\
V47  & 17.134 & 16.783 &0.052234111 &0.05223411&F& 17:32:13.0 &$-$67:02:38.1&1 \\
V48  & 17.048 & --     &0.047920021 &0.04656200&1O&17:31:59.9 & $-$67:03:49.8&1 \\
V64  & 17.086 & 16.656 &0.050162402 &0.05016240&1O&17:31:58.2 & $-$67:03:45.8&2 \\
V72  & 17.560 & 17.180 &0.0436729   &0.0436729 &F& 17:31:29.0 & $-$67:02:33.9&2 \\
SXf1 & 15.075 & 14.673 & --         &0.07949   &-- &17:31:09.8 & $-$66:59:35.6&3 \\
\hline
\end{tabular}
\raggedright
\center{\quad 1: \cite{ole01}, 2:\cite{kal14}, 3: Present work. Star is not a cluster member.}
\end{center}
\end{table*}

\section{The Distance to NGC 6362} 
\label{Distance}

The distance to NGC 6362 has been estimated from four independent approaches as
described in the following sections.

\subsection{From the RR Lyrae stars}

To estimate the distance to NGC 6362 from the RR Lyrae stars we have converted the
individual values of $M_V$ in Table \ref{fisicos} into individual distances and
adopting $E(B-V)=0.063$ derived in $\S$ \ref{RRab_reddening}. Since the values of
$M_V$ come from independent calibrations from RRab and RRc stars we calculated
two independent values of the true distance modulus; 14.499$\pm$0.078 and
14.522$\pm$0.037, corresponding to the distance 7.93$\pm$0.32 and 8.02$\pm$0.15 kpc
from the RRab and the RRc stars respectively.

An independent estimate from $I-$band RR Lyrae P-L relation derived by Catelan
et al. (2004) $M_I = 0.471-1.132~ {\rm log}~P +0.205~ {\rm log}~Z$, with ${\rm log}~Z
= [M/H]-1.765$. We applied these relations to the 34 RR Lyrae stars in Table
\ref{variables}, and found a distance of 7.85$\pm$0.37 kpc.

\subsection{From the SX Phe stars}

\begin{figure}
\includegraphics[scale=0.40]{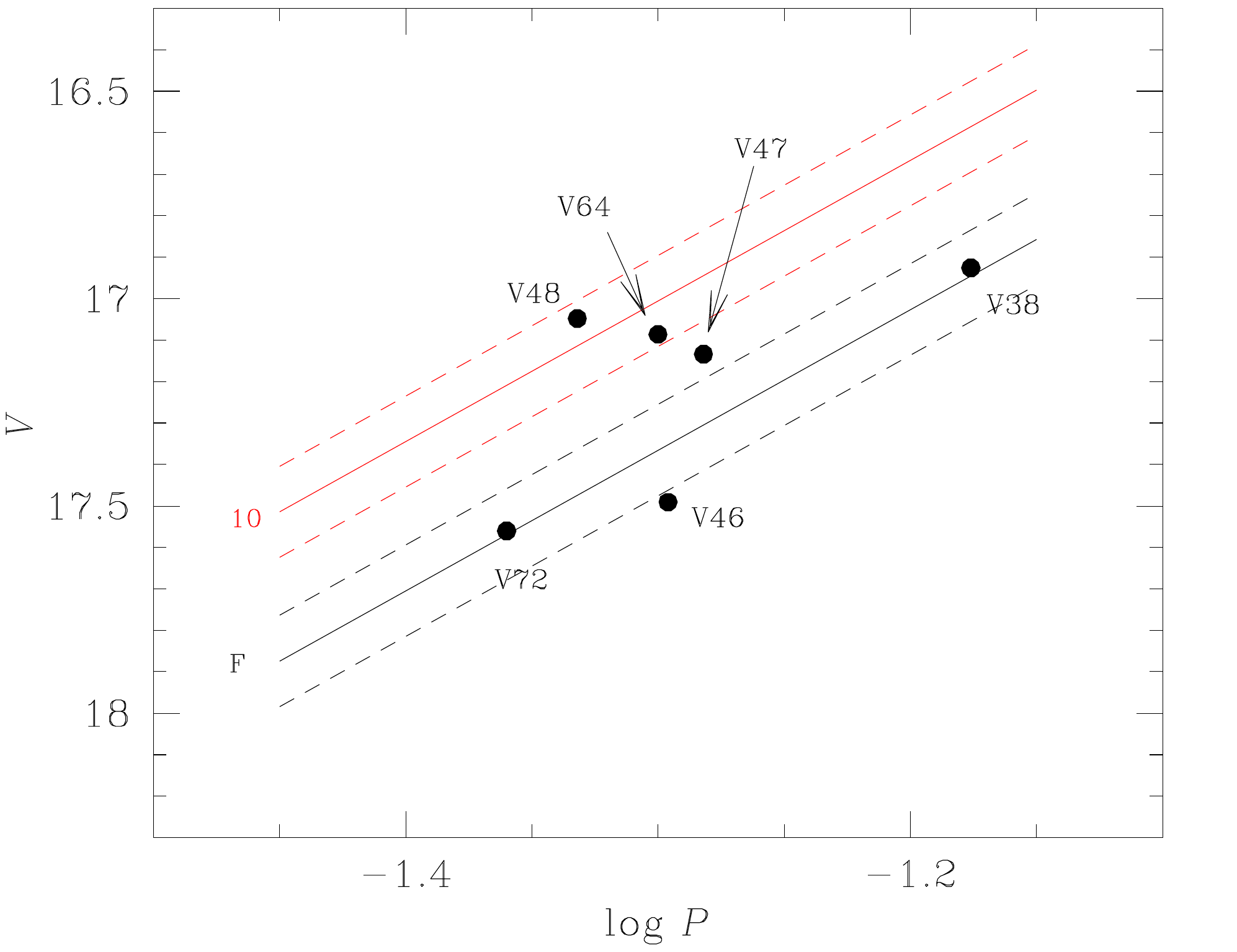}
\caption{The relation between period and luminosity for the SX Phe in NGC 6362. Black 
and red lines correspond to the P-L relation of \cite{CS12} for the fundamental and
first overtone respectively, shifted to a distance of $8.07\pm0.44$ kpc. Segmented
lines correspond to zero-point uncertainty of the P-L relationship. The pulsating mode
assignation to each
star correspond to their position relative mode loci.}
\label{PLSX}
\end{figure}

We have calculated the predicted distance to each SX Phe star, by the P-L relation of \cite{CS12},
taking the periods and pulsating modes listed in Table \ref{SXPhe} and assuming
$E(B-V)$=0.063 for the six cluster members.
The average distance is $8.07\pm0.44$ kpc, in excellent agreement with distance from the RR Lyrae stars.
In Fig.\ref{PLSX} we show the distribution of SX Phe stars in the log P- $V$ plane, where the P-L
relation of \cite{CS12} for the fundamental mode, shifted to the above distance, is displayed.
The corresponding first overtone relation was placed assuming a first overtone to fundamental period
ratio of P1/P0=0.783 \citep{Santo01}.

\subsection{Comments on the tip of the red giant branch method}

The luminosity of the true tip of the RGB can in principle be used to estimate the
distance to globular clusters. The method, originally developed to estimate 
distances to nearby galaxies (Lee et al. 1993), has been used to corroborate the
theoretical constrains that particle-physics may impose on the extension of
the RGB in globular clusters, by comparing the predicted distance by the most
luminous RGB with the results rendered by the RR Lyrae and SX Phe. Consistency
between theoretical predictions and empirical calculations has been found by
for NGC 6229 \cite{Are15}, for M5 \cite{Are16}
and for NGC 6934 \cite{Yep17}. In brief, the bolometric absolute magnitude of
the tip of the RGB was
calibrated by \cite{SaCa97} as:

\begin{equation}
\label{TRGB}
M_{bol}^{tip} = -3.949\, -0.178\, [M/H] + 0.008\, [M/H]^2,
\end{equation}

\noindent
where $[M/H] = \rm{[Fe/H]} - \rm {log} (0.638~f + 0.362)$ and log~f = [$\alpha$/Fe] 
\cite{Sal93}. \cite{Via13} argued that the neutrino
magnetic dipole moment enhances the plasma decay
process, postpones helium ignition in low-mass stars, and extends the red giant branch
in globular clusters. Hence, the true
TRGB may be a bit brighter than the brightest observed stars by an amount between 0.04 and
0.16 mag.

For the case of NGC 6362, applying the largest correction of 0.16 mag, the
method produces an absurdly large distance. We note that the separation between
the Horizontal Branch (HB) and the brightest stars in the RGB in the above mentioned
clusters, is about 3.3 and 4.0 mag in $V$ and $I$ respectively, while in NGC 6362
it is only about 2.3 and 3.5 mag (see also the CMD in \cite{Broc99}). The predicted $M_{bol}^{tip}$ by eq. \ref{TRGB} for
the metallicity and reddening of NGC~6362 is -3.819 mag, and a correction of between
0.5 and 1.0 mag. would be necessary to bring the distance to about 8.0 kpc as
calculated from the RR Lyrae and SX Phe stars. Such correction is way out of the 
theoretical expectations. The method fails for NGC 6362.
We do not really have an explanation for this but, it should be noted that the 
population of the RGB in
this cluster is meager, that those stars at the tip of the RGB could rather be AGB
stars, and that this is a much more metal rich cluster compared with those tested
before.

\section{The horizontal branch of NGC~6362}
\label{sec:HB}

\begin{figure}
\includegraphics[scale=0.40]{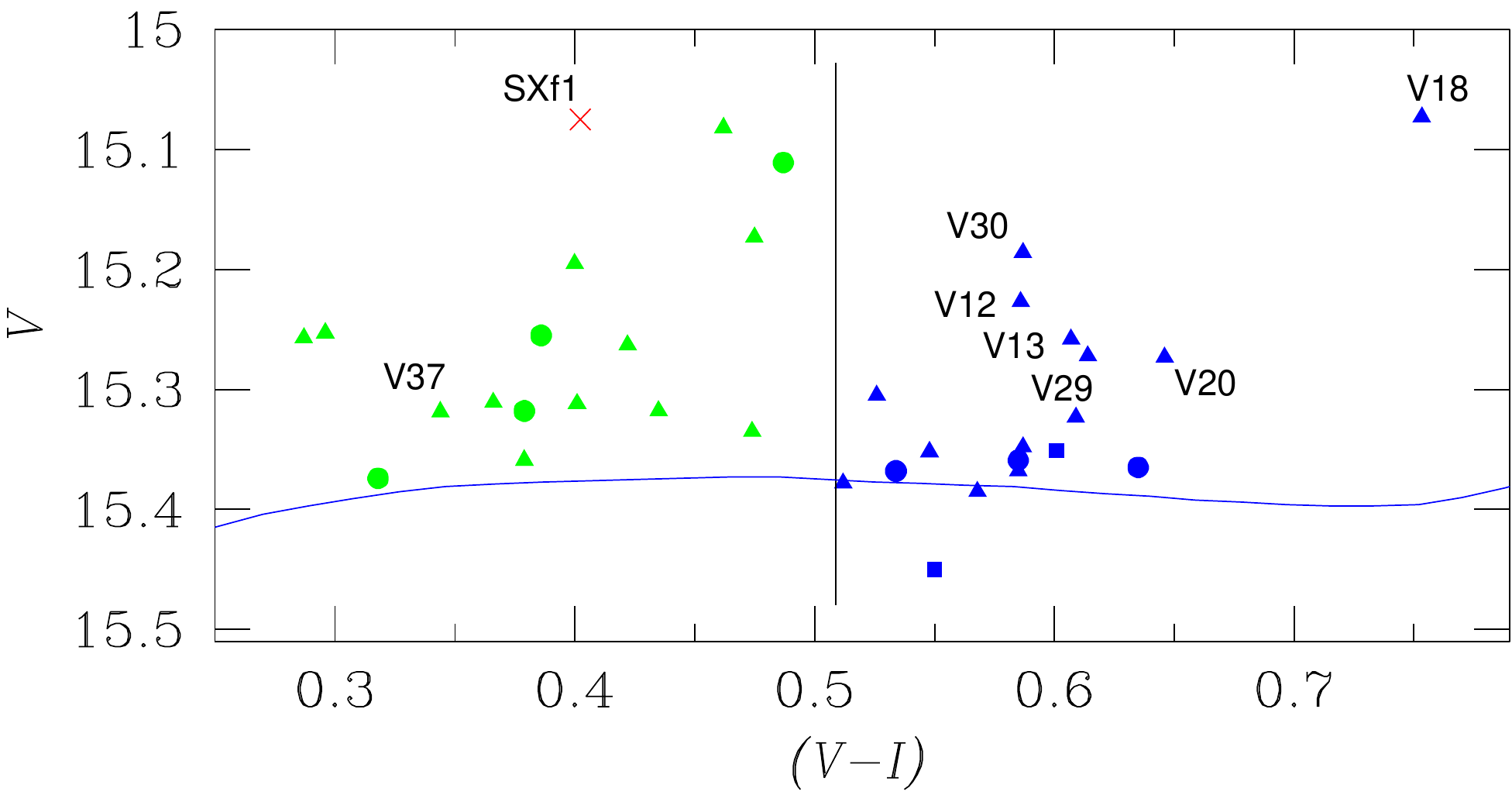}
\caption{The Horizontal Branch of NGC 6362. The stars V12, V13, V20 and V30 are among
the most luminous of the RRab stars and their position in the the log P-Amplitude
diagram (Fig. \ref{figBailey}) suggests they are advanced in their evolution towards the
AGB.}
\label{HB}
\end{figure}

\begin{figure}
\includegraphics[scale=0.49]{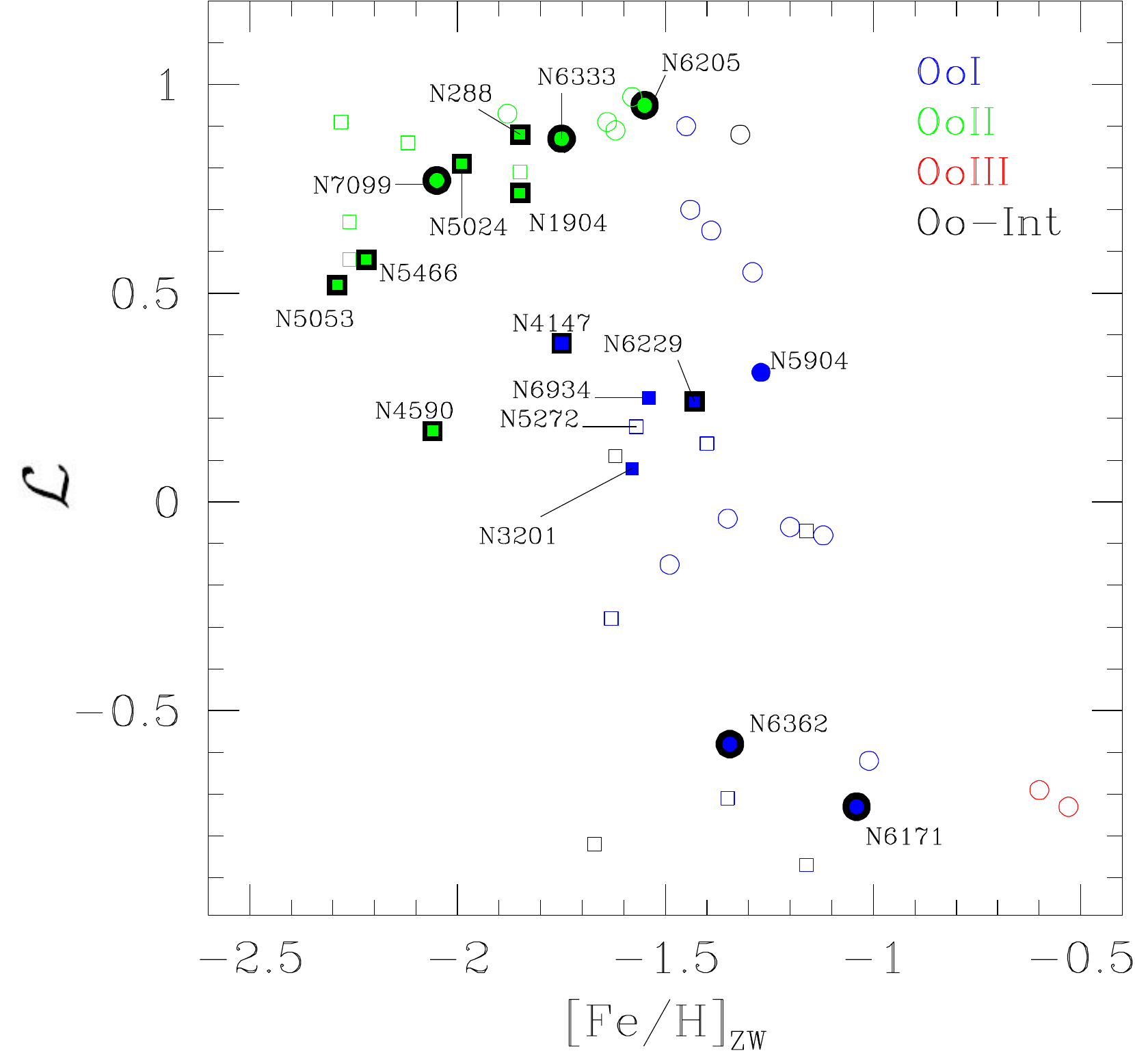}
\caption{The HB structure parameter $\mathcal L$ vs. metallicity. The black-rimmed symbols
represent globular clusters where the fundamental and first overtone modes are well
segregated around the first overtone red edge of the instability strip, as opposed 
to filled non-rimmed symbols. Empty symbols are clusters not studied by our group.}
\label{CATELAN}
\end{figure}

The neat splitting of RRc and RRab stars is clear in Fig. \ref{HB}. This is
significant given that, while the mode splitting has been found in all studied Oo II
type clusters (nine so far) (\cite{Yep17}), it has been observed only some Oo I (three
out of seven so far). Fig. \ref{CATELAN}, which is an updated version of Fig. 16 of
\cite{Yep17}, shows in the HB structure parameter $\mathcal L$ vs [Fe/H]
plane, the segregated clusters (black-rimmed symbols) and the non-segregated ones.
NGC 6362 is now the third OoI cluster where we find the clean mode
segregation, limited by the red edge of the first overtone instability strip. The
other examples are NGC 6229 \cite{Are15} and NGC 6171 (Deras et
al. 2018, in preparation). It has been discussed by \cite{Yep17} that the mode 
segregation versus the mingled mode distribution in
the inter-mode or "either-or" region may be a consequence of the mass loss rates
involved during the He flashes, and hence the resulting mass distribution on the ZAHB,
which in turn controls the mode population of the inter-mode region (Caputo et al.
1978). It is remarkable that all metal-poor and older OoII type clusters do show the
mode segregation.

\section{Summary}

In the present paper we publish new time-series CCD \emph{VI} photometry of the
variable stars in the globular cluster NGC 6362. Via the Fourier decomposition of the
RR Lyrae stars 
we estimated the mean [Fe/H]$_{UVES}$ as $-1.066\pm0.126$ and $-1.08\pm0.16$ and the distance
$7.93\pm0.32$ and $8.02\pm0.15$ kpc from the RRab and RRc stars respectively. 
The reddening $E(B-V)=0.063\pm0.024$ was calculated using the intrinsic colour 
$(V-I)_0$ at minimum colour level for 15 RRab stars. These results
agree very well with previous determinations found in the literature.

Our resulting CMD and the above parameters are in excelent agreement with models 
of the zero-age horizontal branch and the isochrone for 12.1 Gyr from the
Victoria-Regina stellar models of \cite{vdB14}.

Employing the P-L relationship of \cite{CS12} for four fundamental mode and two first 
overtone SX Phe stars, an average distance of $8.07\pm0.44$ kpc was found, consistent
with the results from the RR Lyrae stars.

We found a new SX Phe in the field of our images that seems to be a foreground star.

The distribution of the RRL stars in the HB and the Period-Amplitude diagram for
this 
cluster reveal the presence of four RRab stars (V12, V13, V20 and V30) likely to be 
advanced in their evolution towards the AGB. If this is the case, as they evolve towards 
the red, they should display a positive secular period variation; a study yet to be performed.
The fundamental and first overtone RRL stars, independently of whether they have
Blazhko 
modulations and/or two excited pulsation modes, are neatly segregated around the red
edge of the first overtone instability strip, which is a characteristic in OoII type
clusters but only in some of the OoI clusters studied thus far.

\acknowledgements
We are grateful to Dr. Daniel Bramich for his DanDIA software and for guiding our data
reduction process. AAF recognizes and thanks the support of UNAM via the DGAPA
project IN104917. JAA wishes to thank the Instituto de Astronom\'{\i}a of the
Universidad Nacional Aut\'{o}noma de Mexico for hospitality. We have made an extensive
use of the SIMBAD and ADS services, for
which we are thankful.


%
\bibliographystyle{an}
\bibliography{6362_biblio}
%

\appendix

\section{Comments on individual variables}
\label{individual}

\subsection{V18}
This star presents amplitude modulations and has been considered as a Blazhko
variable by \cite{ole01} and Smo17. It appears as the most luminous and reddest
RRab in the CMD of Fig. \ref{CMD_6362}. The light curve in the $I$-band is of low
amplitude and anomalously bright, hence the red position of the star, but no
suggestion of advanced evolution is found from its position in the log P- Amplitude
diagram. 

\subsection{V37} 
\begin{figure}
\includegraphics[scale=0.42]{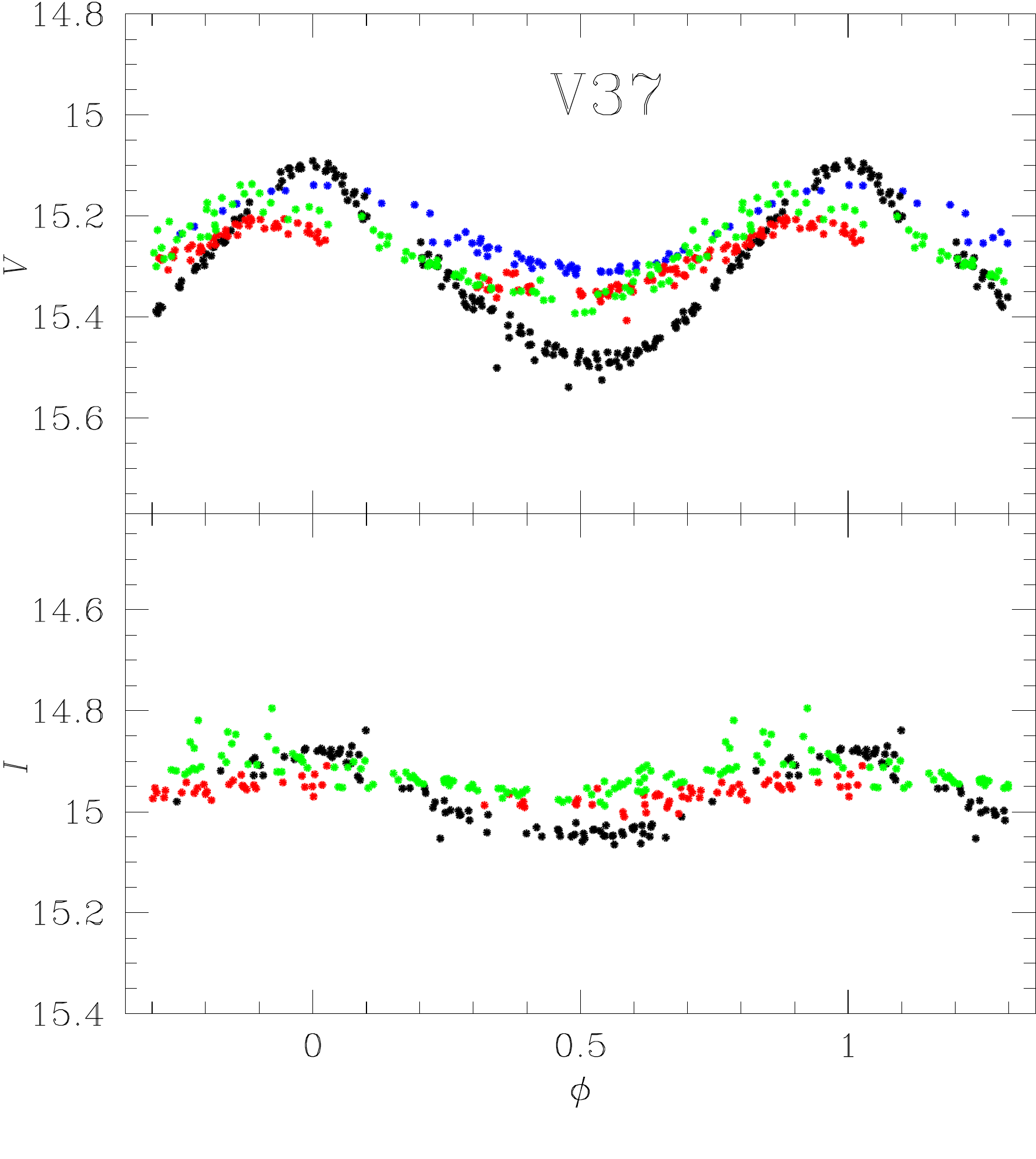}
\caption{V37 light curve. Colors are as in Fig. \ref{RRab}. Colours are as in Fig.
\ref{RRab}.}
\label{V37}
\end{figure}

\begin{figure}
\includegraphics[scale=0.85]{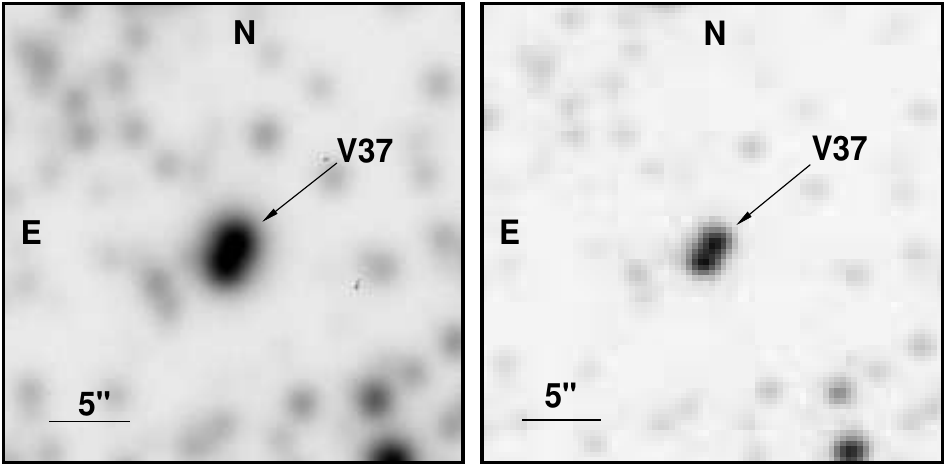}
\caption{V37 charts in the reference images from CASLEO (left) and from SWOPE (right).}
\label{V37chart}
\end{figure}

This variable was identified by \cite{ole01} as a double mode pulsator
with one mode being non-radial. The star has been reconsidered by Smo17
and they identified the two modes, being at least one of them non-radial.
On the basis of
the light curve shape and amplitudes of the two modes, they postulate
that the star is not a typical RRc but a beating variable of a new type. The light
curve from our data is shown in Fig. \ref{V37} and, although less dense, it is similar
and consistent with the light curves shown by \cite{ole01} and Smo17. We would
like to note however, that the star is not isolated but that a very close
neighbour is
present at only 2.65 arc seconds to the SE from V37. In Fig. \ref{V37chart} we show
this pair in the reference images of our CASLEO observations, of rather high seeing,
and SWOPE with much better seeing conditions. Even under better conditions, in the
SWOPE detector it is difficult to isolate and measure the two stars individually.
Thus, contamination of V37 from the flux of the neighbour is most likely present in
the photometry of previous and present studies.
We have examined all differential images by blinking them along with the reference
images and found that the star at the NW, i.e. that labelled V37, is variable and
got some hint that star at the SE may also exhibit some variations, although this is
not a firm conclusion given the blending conditions of the pair. If this proves to be
true, then at least part of the observed modulations might be an artifact. A
confirmation of the double mode nature and properties of V37 with a wider scale
telescope is highly desired.

In the CMD (e.g. Fig. \ref{HB}), V37 is near the hot edge of the first overtone
instability strip although its colour has certainly been altered by the presence of
the close neighbour. In the Log P -Amplitude diagram it mingles well among the RRc
stars and falls, as expected, near the predicted locus for the first overtones. 

\subsection{SXf1}

\begin{figure}
\includegraphics[scale=0.42]{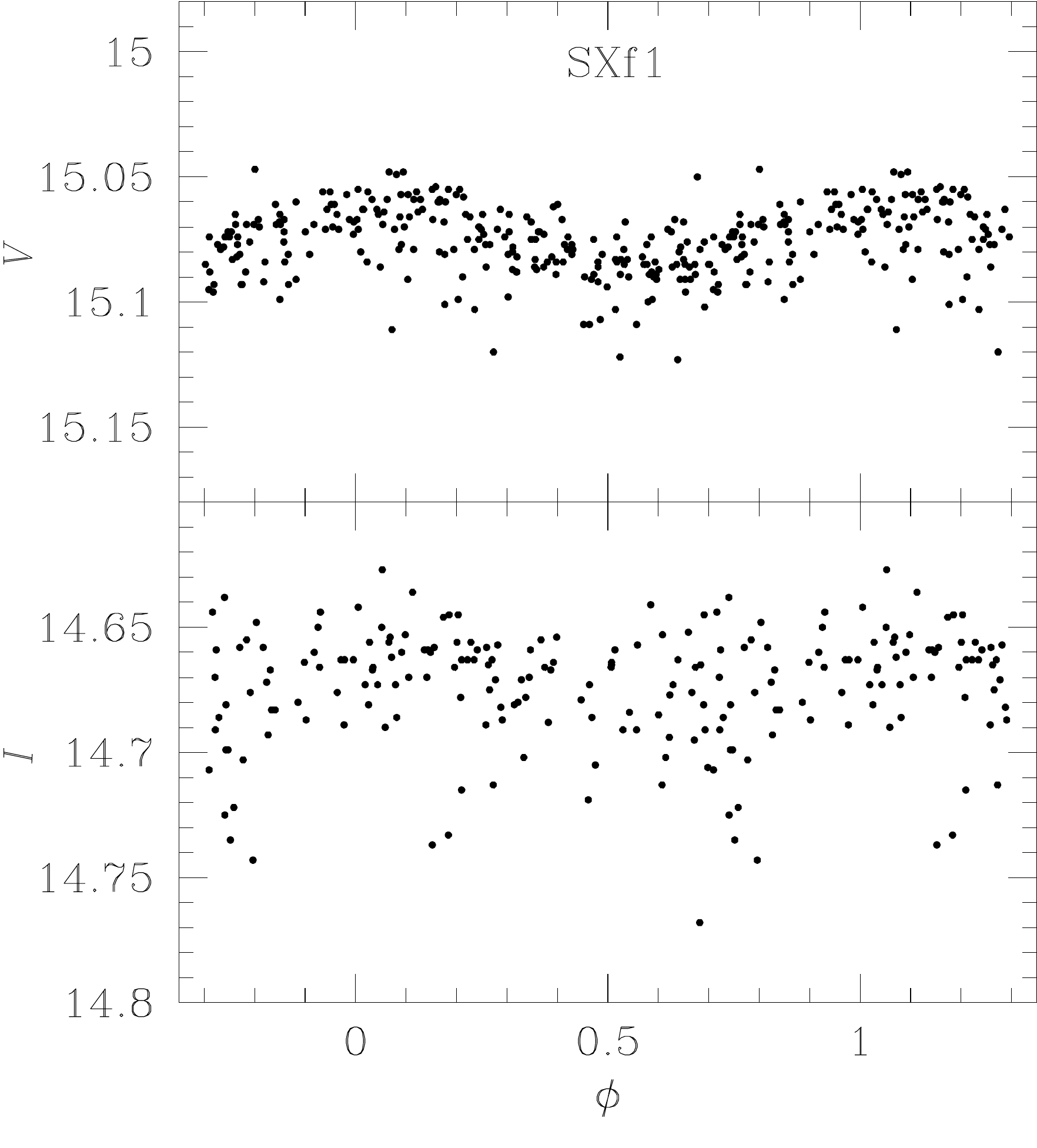}
\caption{The newly identified SX Phe in the field of NGC 6362. The star is not a
cluster member but a foreground object. The light curves are phased with a period of
0.07949 d.}
\label{SXf1}
\end{figure}

The light curve of this star shows clear sinusoidal variations (Fig. \ref{SXf1}) of 
small amplitude and a period of 0.07949 d, typical of SX Phe stars. The star lies at
the HB level, i.e. some 2 magnitudes brighter, although of similar colour, than the SX
Phe cluster members. Therefore, we classify it as a foreground field SX Phe star.

\section{Identification chart}
\label{chart}

\begin{figure*}
\includegraphics[scale=0.17]{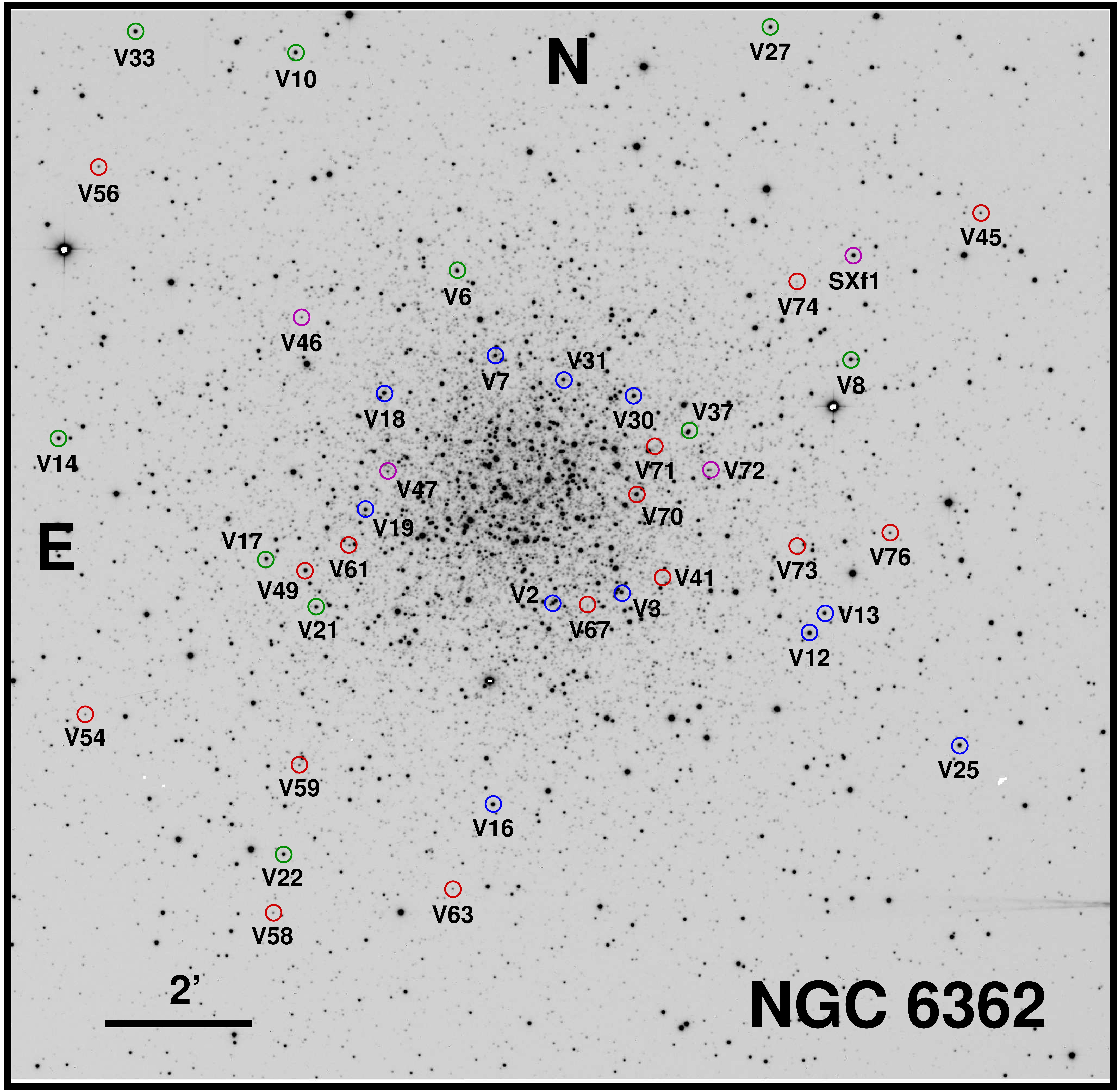}
\caption{Variable stars in the external field of NGC 6362. The image is the $V$
reference image
employed for the differential imaging in the data from the SWOPE telescope. The field
is about 14.5$\times$14.5 arcmin$^2$. The newly identified foreground SX Phe star is
labelled as SXf1. Colours are used for different types of variables: blue for RRab
stars, green for RRc, purple for SX Phe and red for eclipsing binaries.}
\label{chartA}
\end{figure*}

\begin{figure*}
\includegraphics[scale=0.17]{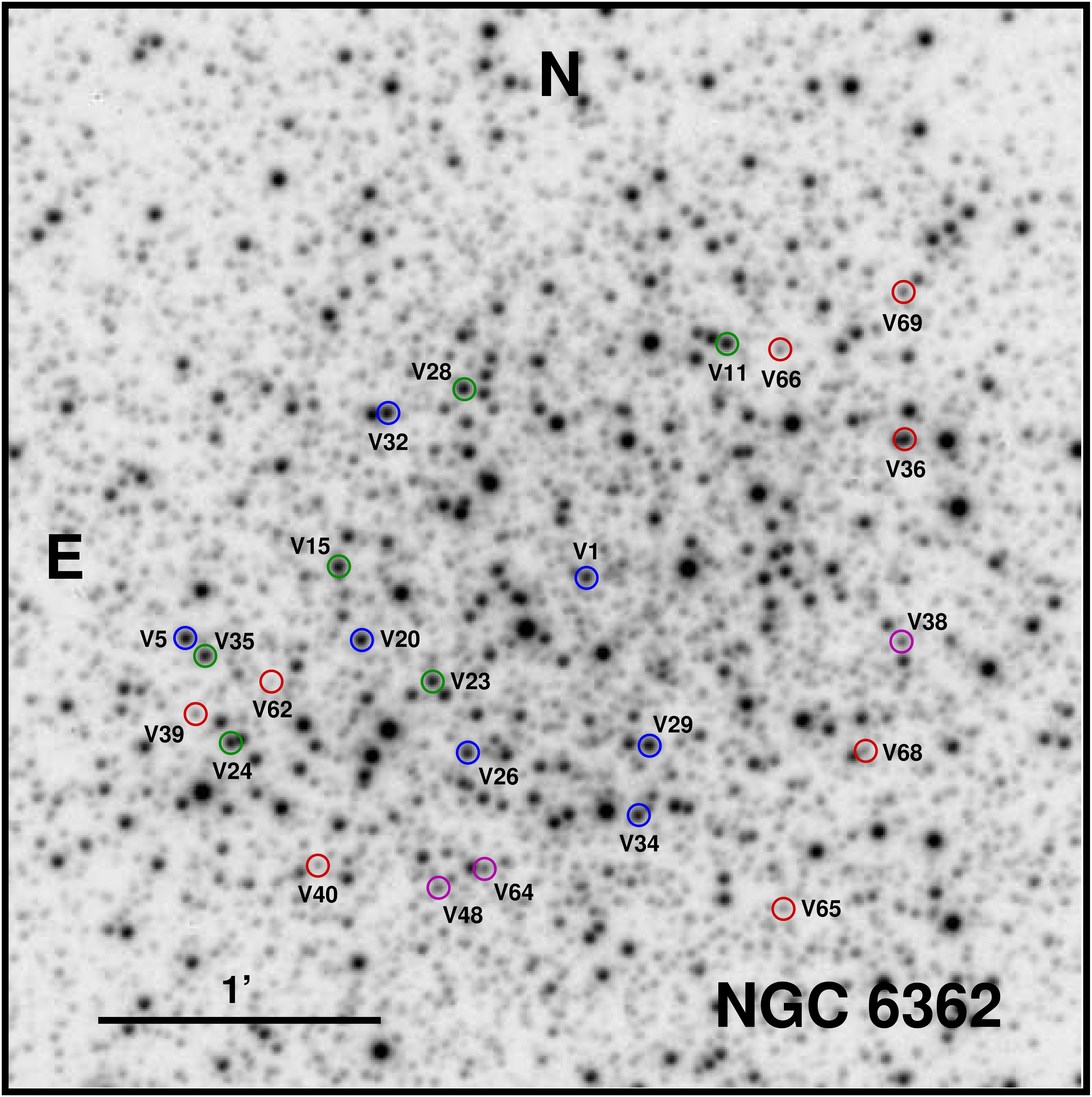}
\caption{Variable stars in the core region of NGC 6362. The field
is about 3.75$\times$3.75 arcmin$^2$. The colour code is as in
Fig. \ref{chartA}.}
\label{chartB}
\end{figure*}

All the known variables in the field of our images are identified in charts
\ref{chartA} and \ref{chartB}.
\cite{kal14} provided identification
charts and RA and Dec coordinates, for all the variables stars in their study.
We noted however that some coordinates did not correspond to their
identification. M.~Rozyczka (private comm.)
pointed out that the coordinates of V42, V45, V49, V53--V58, V60, V63, and V75--V77
are in fact incorrect in their Table~1, and kindly provided the correct coordinates.
Aiming to clarify the identifications of these stars, in Table \ref{ARD_corr} we list
the correct coordinates, which are consistent with the
identifications in the charts of Fig. \ref{chartA} and \ref{chartB}.

\begin{table*}
\begin{center}
\caption{Corrected coordinates of binary stars in NGC 6362. The remainder of the
binary stars are correctly identified in Table 1 of \cite{kal14}.}
\label{ARD_corr}

\begin{tabular}{lllll}
\hline
Variable &  RA & DEC   & RA &  DEC  \\
Star ID  &  (decimal) & (decimal)   & (h m s)  &  ($^\circ~'~''$) \\
&&&&\\
\hline
V42& 262.78754& $-$66.860809&17:31:09.0 &--66:51:38.9\\
V45& 262.71952& $-$66.983002&17:30:52.6 &--66:58:58.8\\
V49& 263.10043& $-$67.066788&17:32:24.1 &--67:04:00.4\\
V53& 263.29060& $-$66.856061&17:33:09.7 &--66:51:21.8 \\
V54& 263.22510& $-$67.098769&17:32:54.0 &--67:05:55.5 \\
V55& 263.22533& $-$66.926742&17:32:54.0 &--66:55:36.2 \\
V56& 263.21986& $-$66.974452&17:32:52.7 &--66:58:28.0 \\
V57& 263.19214& $-$66.925905&17:32:46.1 &--66:55:33.2 \\
V58& 263.11699& $-$67.145429&17:32:28.0 &--67:09:43.5 \\
V60& 263.09729& $-$66.924169&17:32:23:3 &--66:55:27.0 \\
V63& 263.01446& $-$67.139542&17:32:03.5 &--67:08:22.4 \\
V75& 262.80971& $-$66.924412&17:31:14.3 &--66:55:27.8 \\
V76& 262.76813& $-$67.056661&17:31:04.4 &--67:03:24:0\\
V77& 262.71325& $-$66.924622&17:30:51.1 &--66:55:28.6 \\
\hline
\end{tabular}
\end{center}
\end{table*}

\end{document}